# Generation of tunable ultrashort pulses in the near infrared

C. Fourcade Dutin [1] and D. Bigourd [1]

[1] Institut FEMTO-ST, Département Optique, UMR 6174, Université Bourgogne Franche-Comté-CNRS, 25030 Besançon, France
E-mail: damien.bigourd@femto-st.fr

**Summary:** A laser source delivering ultrashort pulses (50-100 fs) tunable from 820 nm to 1200 nm has been developed. It is based on the filtering of a continuum in the Fourier plane of a zero dispersion line without a phase compensator. We have also numerically investigated the impact of the residual spectral phase in order to guarantee ultrashort pulses.

**Keywords:** Ultra-short pulse generation, near infrared, continuum

## 1. Introduction

The generation of near infrared tunable ultrashort pulses from an oscillator at fixed center wavelength is of prime interest for many applications as nonlinear spectroscopy [1] or synchronization of several laser pulses with low timing jitter or drift [2]. In this latter case, the synchronization without complex electronic device can be achieved when a unique pulse generates the other one through nonlinear processes in a photonic crystal fiber (PCF). For example, the pump and signal in an optical parametric amplifier are often generated from a common oscillator [2]. However, the dispersion feature of the PCF defines the characteristic of the generated continuum. Pumping the PCF in the anomalous dispersion regime near a single zero dispersion wavelength (ZDW) leads often to amplitude and timing jitter of the pulse. Alternatively, all normal dispersion (ANDI) fiber can be exploited to generate a very large bandwidth with smooth spectral intensity and phase [3]. In this case, the continuum is considered as highly coherent since the nonlinear process is dominated by self-phase modulation (SPM) and optical wave breaking (OWB) [4], and thus can be used to generate few cycle pulses [5]. Alternatively, tunable ultrashort pulses can be produced when an adjustable part of the continuum is selected with bandpass filters and the power [6]. In this paper, we focus the investigation to the selection of tunable near infrared sub-100 fs pulses by filtering the broad bandwidth with only one zero dispersion line without phase compensator. We have numerically demonstrated that the uncompensated spectral phase deteriorates marginally the pulse duration even at maximum power. We have experimentally achieved the production of sub-100 fs pulses tunable from 800 nm to 1200 nm.

## 2. Simulation

Numerical simulations have been conducted by integrating the nonlinear Schrödinger equation along the propagation describing the evolution of the slowly varying total electric field in the ANDI fiber. The equation has been solved with the standard split-step Fourier with the fiber parameters. The experimental conditions have been used in the simulation. The pulse has a duration of 80 fs at full width half maximum (FWHM) and a center wavelength of 1030 nm. The maximum average power is 500 mW for a repetition rate of 76 MHz. Fig.1.a displays the generated continuum in the 7 cm long fiber for ~410 mW. The bandwidth increases mainly due to SPM and OWB [3]. It extends from 750 to 1300 nm. The continuum has been narrowed with several bandpass filters centered at 900 and 1100 nm with a bandwidth equals to 5.6 THz (Fig.1.a-blue and red lines).

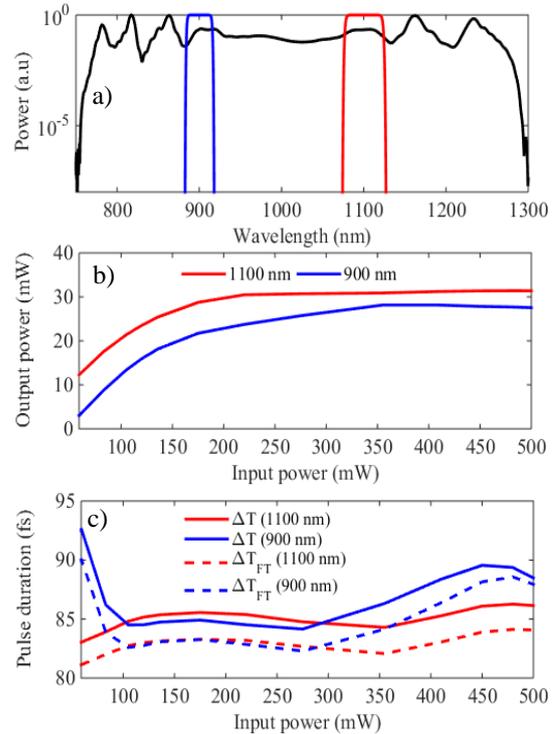

**Fig. 1.** a.) Simulated continuum (black line) at 410 mW. The red and blue lines correspond to the bandpass filters. b) Output power when the filter is centered at 900 nm and 1100 nm with a bandwidth of 5.6 THz. c) Durations of the filtered pulses ($\Delta T$-solid lines) and the Fourier transform limited pulses ($\Delta T_{FT}$-dashed lines).

When the input power increases, the output power of the filtered continuum is enhanced (Fig.1.b) and saturates around 30 mW from an input power of ~200 mW. Simultaneously, the pulse duration is approximatively constant around 85 fs near the Fourier transform limit (Fig.1.c). These two characteristics are very important to reach higher peak power in the ultrashort time scale. The slight variation of the pulse duration with the power is due to the modification of the continuum structure.

In order to reach shorter pulses, we have investigated the impact of the filter bandwidth (at FWHM) on the duration (Fig.2). As expected, the Fourier transform limited pulse duration decreases with a larger spectrum (dahed lines-Fig.2). However, the contribution of the spectral phase becomes more pronounced and therefore the pulse duration increases from its limit (solid lines-Fig.2). From a bandwidth equals to ~28 nm at 900 nm, the pulse duration increases suddenly due to the local variation of spectral phase near the dip in the continuum (at 884 nm in Fig.1.a). At 1100 nm, the increase of the pulse duration is smoother since the filter is not located near a dip in the continuum.

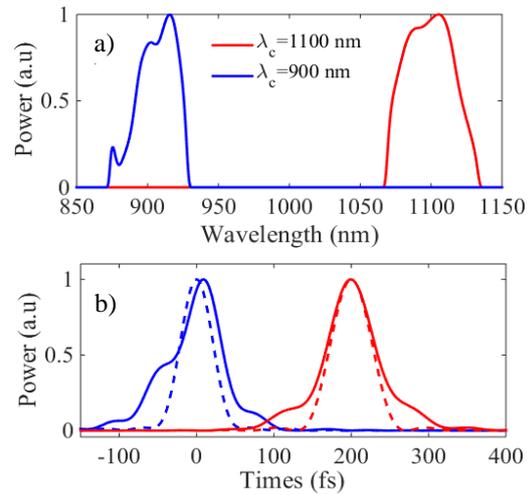

**Fig. 3.** a) Spectrum of the filtered continuum at $\lambda_c$=1100 nm and $\lambda_c$=900 nm. b) Corresponding pulse shape. The dashed lines represent the trace of the Fourier transform limited pulses

## 3. Experiment

### 3.1. Experimental set-up

The experimental set-up is shown in Fig.4.a. The oscillator delivers sub-80 fs pulses centered at 1030 nm at 76 MHz. The beam is injected in the ANDI fiber and the maximum bandwidth of the continuum is achieved with 410 mW. It extends from 720 nm to 1300 nm (Fig.4.b).

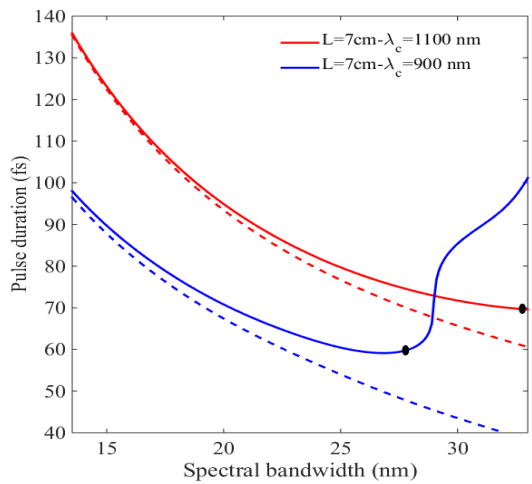

**Fig. 2.** Pulse duration as a function of the spectral bandwidth (FWHM) for $\lambda_c$=1100 nm and $\lambda_c$=900 nm. P=410 mW. The dashed lines correpond to the duration of the Fourier transform limited pulses.

In the experiment, we will target the bandwidth that allow the minimum pulse duration without any phase compensation (black circles-Fig.2). In this case, the filtered spectra and the pulse shapes are shown in Fig. 3. The minimum pulse duration is 60 fs and 70 fs for $\lambda_c$=1100 nm and $\lambda_c$=900 nm, respectively. The impact of the phase can be minimized for shorter fiber since the whole spectral phase should be decreased. A detailed work is currently under investigation.

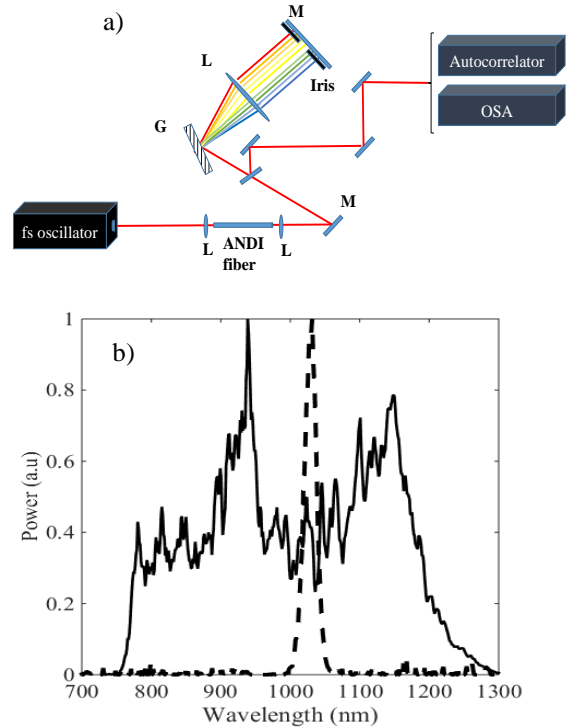

Fig. 4. a) Experimental set-up. M. Mirror, L. Lens b) Continuum generated in the ANDI fiber with 410 mW (solid line) and spectrum of the oscillator (dashed line).

Firstly, we performed a preliminary test in order to verify the possibility to obtain an ultrashort pulse at a selected wavelength. The continuum has been narrowed by a low order super-Gaussian bandpass filter centered at 905 nm with a bandwidth equal to 25 nm at FWHM without any compressor. The filtered spectrum is shown in Fig.5.a (black dashed line). The modulation are weak within this spectral bandwidth. The autocorrelation trace has a Gaussian shape with a pulse duration of 140 fs (Fig.5.b-black solid line) while the Fourier transform limited pulse duration is 80 fs (Fig.5.b-black dashed line). Assuming a Gaussian profile (Fig.5.b-green dotted line), we conclude that it is possible to generate a 100 fs pulse with some residual spectral phase when the spectrum is filtered in the continuum.

For the longer scan, the grating is rotated and the fine tuning is achieved with the iris. In order to estimate the amount of spectral phase, the autocorrelation traces of the Fourier transform limited pulses are also plotted in Fig.6.b (dashed lines). When the central wavelength varies from 820 nm to 1200 nm, the pulse duration of the autocorrelation trace ranges from 57 fs to 113 fs near the Fourier transform limit. The pedestals are mainly due to the spectral phase and the contribution from the modulated spectrum is relatively weak since the Fourier transform limited pulses have bell-shaped profiles. This spectral phase can be compensated for example with chirped mirrors [5], but they need to be specifically adjusted for a specific range of wavelength.

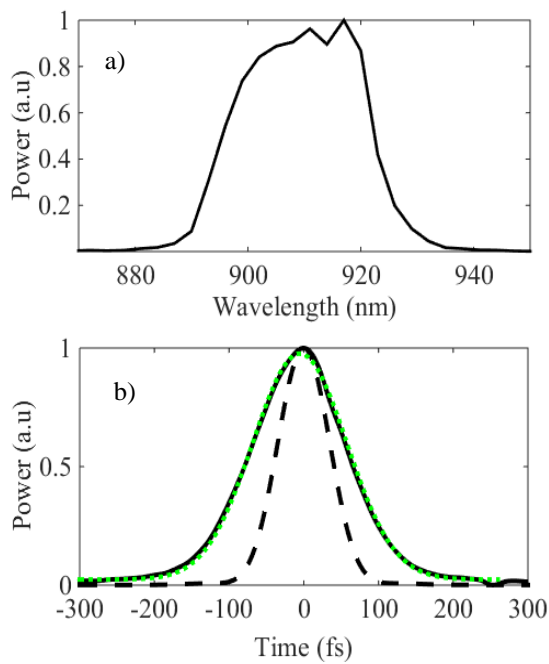

**Fig. 5.** a) Spectrum selected by a bandpass filter centered at $\lambda_c$=905 nm. b) Corresponding autocorrelation trace. The green line corresponds to a fit with a Gaussian profile. The dashed line represents the trace of the Fourier transform limited pulses.

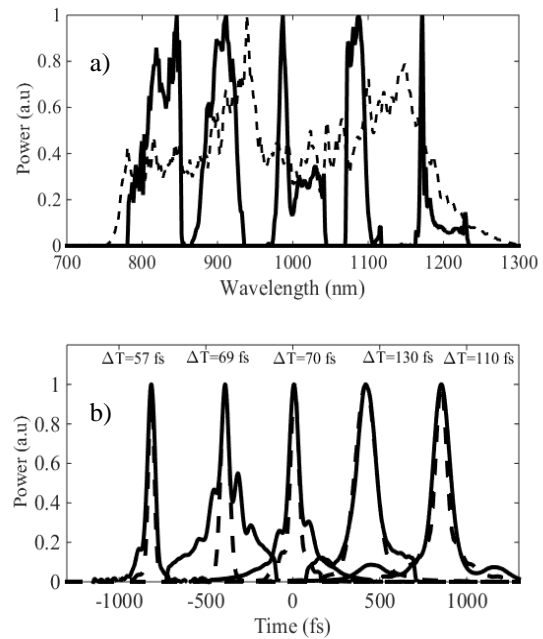

**Fig. 6.** a) Selection of filtered spectrum in the Fourier Plane. The dashed line corresponds to the continuum. b) Corresponding autocorrelation trace (solid lines). The dashed lines represent the trace of the Fourier transform limited pulses.

In order to achieve a tunable filtering system, we develop a folded 4-f zero dispersion line composed of a grating with 1200 lines/mm, a lens (f=100 mm) and a flat mirror (Fig. 4). The grating and the folded mirror are placed on translation stages to optimize the dispersion induced by the lens at each wavelength.

**3.2 Results**

Some filtered spectra and corresponding autocorrelation traces are shown in Fig.6 for the total power of 410 mW. For a short scan of the wavelength, an iris can be translated in front of the folded mirror.

## 4. Conclusions

We demonstrated the possibility to generate tunable 50-100 fs pulses from 800 nm to 1200 nm by selecting a part of a continuum with a zero dispersion line. The uncompensated spectral phase deteriorates slightly the pulse duration. The power of the generated ultrashort pulses is relatively low (1-30 mW) mainly due to the filtering and the transmission of the zero dispersion line. However, this level is sufficient to seed a fiber optical parametric amplifier [7]. In a future work, the generated ultra-short pulse (the signal) and a pulse from the oscillator (the pump) will be stretched to few tens of picosecond to decrease the peak power. The pump pulse will be amplified in several fiber doped fiber amplifiers to increase the peak power. The two pulses will be injected in a photonic crystal fiber to perform the parametric process enabling ultra-short pulse amplification with very large bandwidth [8].

## Acknowledgements

This work is supported by the ANR (LABEX Action ANR-11-LABS-0001-01 and FiberAmp projects ANR-16-CE24-0009) and the conseil Régional de Franche-Comté.